# Pyroelectric and electrocaloric properties of core-shell $Hf_xZr_{1-x}O_2$ nanoparticles: theory and experiment

Anna N. Morozovska[1*], Oleksandr S. Pylypchuk[1], Nicholas V. Morozovsky[1], Eugene A. Eliseev[2†], and Dean R. Evans[3‡]

[1] Institute of Physics of the National Academy of Sciences of Ukraine,
46, Nauky Avenue, 03028 Kyiv, Ukraine

[2] Frantsevich Institute for Problems in Materials Science of the National Academy of Sciences of Ukraine,
3, str. Omeliana Pritsaka, 03142 Kyiv, Ukraine

[3]Zone 5 Technologies, Special Projects, San Luis Obispo CA 93401, USA

**Abstract**

Nanosized hafnia-zirconia ($Hf_xZr_{1-x}O_2$) in the form of thin films, multilayers, and nanoparticles is one of the most promising CMOS-compatible ferroelectric materials for advanced electronic memories and logic devices. Using the Landau-Ginzburg-Devonshire free energy functional with trilinear and biquadratic couplings of polar, nonpolar, and antipolar order parameters, we analyze the pyroelectric and electrocaloric properties in an ensemble of spherical core-shell $Hf_xZr_{1-x}O_2$ nanoparticles. To test the theoretical model, we experimentally measured the temperature dependence of the electric charge accumulated in pressed powders consisting of oxygen-deficient core-shell $Hf_{0.5}Zr_{0.5}O_2$ nanoparticles with an average size of 7 nm. The observed temperature-dependent behavior of the accumulated charge and its derivative are in qualitative agreement with the calculated polarization and pyroelectric coefficient for the ensemble of densely packed spherical core-shell $Hf_xZr_{1-x}O_2$ nanoparticles; this suggests that the theoretical model captures the physical mechanisms responsible for the experimentally observed charge accumulation. The combined theoretical and experimental results provide a physical foundation for the future development of CMOS-compatible $Hf_xZr_{1-x}O_2$ nanoparticles for pyroelectric and electrocaloric applications.

---

*Corresponding author, e-mail anna.n.morozovska@gmail.com

† Corresponding author, e-mail: eugene.e.eliseev@gmail.com

‡ Corresponding author, e-mail: deanevans@zone5tech.com

## 1. Introduction

Nanosized hafnia-zirconia $Hf_xZr_{1-x}O_2$ ($0 \leq x \leq 1$) in the form of thin films [1], multilayers [2], and superlattices [3] is recognized as one of the most promising CMOS-compatible ferroelectric materials for advanced electronic memories [4] and nanoelectronics [5], including steep-slope and/or negative capacitance field-effect transistors [6, 7]. In particular, Cheema et al. [8, 9] experimentally studied polar properties and charge storage in ferroelectric-antiferroelectric $HfO_2/ZrO_2$ multilayers and superlattices deposited on a silicon substrate and demonstrated a pronounced enhancement of capacitance due to the negative capacitance state of $HfO_2$ layers.

Despite promising applied aspects, the physical nature of the emergence of ferroelectricity in nanoscale hafnia-zirconia is not established [10, 11, 12]. According to recent theoretical studies, it may have an extrinsic nature [13, 14] related to finite size effects [15, 16]. These and many other theoretical calculations seek to explain well-established experimental results, which reveal that the ferroelectric (ferrielectric or antiferroelectric) long-range order emerges in $Hf_xZr_{1-x}O_2$ thin films only when their thickness is reduced to the nanoscale (less than 20 – 30 nm) [17, 18, 19]. The origin of ferroelectricity in $Hf_xZr_{1-x}O_2$ may be related to the trilinear couplings of polar, nonpolar, and antipolar order parameters [20, 21, 22]; these couplings lead to the size-induced transition from the bulk non-polar monoclinic m-phase (space group $P2_1/c$) to the polar orthorhombic o-III phase (space group $Pca2_1$). The film-substrate lattice mismatch stabilizes the o-III phase in $Hf_xZr_{1-x}O_2$ thin films [16], whereas in $Hf_xZr_{1-x}O_2$ nanoparticles, chemical strains created by surface ions and/or oxygen vacancies generate stresses that stabilize the ferroelectric phase [14, 23]. X-ray diffraction (XRD) studies [24, 25, 26] reveal a mixture of o-phases in small (3 – 30 nm) $Hf_xZr_{1-x}O_2$ nanoparticles with x = 1, 0.6, 0.5, and 0.4, which may include the nonpolar o-I (space group *Pbca*), antipolar o-II (space group *Pbcm*), and polar o-III (space group $Pca2_1$) phases, but these studies cannot distinguish between the phases. The orthorhombic phases are stabilized in small $Hf_{0.5}Zr_{0.5}O_2$ nanoparticles annealed in $CO+CO_2$ atmosphere [27].

The direct relation of pyroelectric and electrocaloric properties with the ferroelectric state are studied for $Hf_xZr_{1-x}O_2$ thin films. The emergence of ferroelectricity in $Hf_xZr_{1-x}O_2$ thin films, as well as its temperature and chemical stability, depend strongly on Hf content [17], dopants [28, 29], surface and/or interface effects [17, 30]. These studies demonstrate that the pyroelectric and electrocaloric properties of $Hf_xZr_{1-x}O_2$ can be tailored through composition and defect engineering. In particular, acceptor doping can strongly influence the phase transitions of $HfO_2$ thin films, making them attractive for energy-related pyroelectric and electrocaloric applications [31]. Doping with silicon or aluminum can have a strong influence on the pyroelectric and electrocaloric properties of $HfO_2$ thin films [32, 33]. Furthermore, $Hf_xZr_{1-x}O_2$ films exhibit a giant electrocaloric effect [34], whose sign can be controlled by tuning the Hf content x:

a giant positive electrocaloric effect was observed in $Hf_{0.2}Zr_{0.8}O_2$ thin films, whereas a giant negative electrocaloric effect was observed in $Hf_{0.5}Zr_{0.5}O_2$ thin films.

To the best of our knowledge, there are no direct experimental observations of ferroelectric (or ferroelectric-like) polarization hysteresis loops in $Hf_xZr_{1-x}O_2$ nanoparticles obtained using techniques such as piezoresponse force microscopy. Likewise, direct observations of their pyroelectric and/or electrocaloric properties have not been reported. However, a pronounced charge accumulation effect was measured in $Hf_{0.5}Zr_{0.5}O_2$ nanopowders sintered by the auto-combustion sol-gel method [35]. The effect may indicate the presence of ferroelectric-like polar regions in the nanoparticles. A colossal dielectric response of small (5 – 10 nm) oxygen-deficient $Hf_xZr_{1-x}O_2$ nanoparticles prepared by the solid-state organonitrate synthesis and annealed in a $CO+CO_2$ atmosphere has been observed over the temperature range 38 – 88°C [36]. This was explained by the complex interplay between a possible diffuse ferroelectric-paraelectric phase transition and the Maxwell-Wagner-Sillars (MWS) effect, the latter arises from the formation of volume charges at interfaces between different materials, such as with the ferroelectric nanoparticle-air interface considered in Ref. [36]. Indeed, as shown in Refs. [37, 38, 39], the MWS-type effective dielectric permittivity reaches colossal values (greater than $10^4$) at low frequencies. Mesoscopic inhomogeneities in electrical conductivity (mainly between grains/particles and their boundaries) give rise to interfacial charge accumulation, producing MWS-type polarization and internal barrier layer capacitance (IBLC), resulting in the experimentally observed colossal effective permittivity in nanograined ferroelectric ceramics, nanopowders, and/or nanocomposites [40]. Additional mechanisms, including nonpolar shells with very high lattice strain gradients (LSG) [41] and polaronic effects [42], have also been proposed to contribute to colossal permittivity. Additional contributions to the colossal permittivity may arise from inhomogeneous layers between the electrodes and the sample, known as surface barrier layer capacitance (SBLC). Both the IBLC and SBLC effects produce a colossal dielectric response at low frequencies. These interfacial polarization mechanisms can be described by the effective medium approach (EMA), as discussed in Refs. [43, 44], and in the works of Petzelt et al. [45] and Richetsky et al. [46]. For systems where LSG and/or polaronic effects dominate, the dielectric response is weakly frequency-independent up to the MHz frequency range. In addition to the dielectric phenomena discussed above, several other experimental observations have been reported in $Hf_xZr_{1-x}O_{2-y}$ nanoparticles prepared by the solid-state organonitrate synthesis [47]. Superparamagnetic-like and superparaelectric-like responses were observed, together with the posistor effect and significant charge accumulation [47].

In this work, we analyze the pyroelectric and electrocaloric properties of an ensemble of spherical core-shell $Hf_xZr_{1-x}O_2$ nanoparticles with randomly oriented crystallographic axes. We use the Landau-Ginzburg-Devonshire (LGD) free energy functional with higher power terms and trilinear and biquadratic couplings of the polar ($Q_{\Gamma 3}$), antipolar ($Q_{Y4}$) and nonpolar ($Q_{Y2}$) order parameters. These order parameters are the dimensionless amplitudes of the corresponding polar ($\Gamma_{3-}$), antipolar ($Y_{4-}$), and nonpolar ($Y_{2+}$) phonon modes. To complement the theoretical calculations, we measured the temperature-dependence of the accumulated electric charge in pressed powders consisting of oxygen-deficient $Hf_{0.5}Zr_{0.5}O_2$ nanoparticles with an average size of 7 nm. Analysis of these experimental results using the EMA, followed by comparison with the LGD theoretical predictions, suggests that oxygen-deficient $Hf_xZr_{1-x}O_2$ nanoparticles exhibit ferroelectric, pyroelectric, and electrocaloric behavior.

## 2. Calculations of pyroelectric and electrocaloric properties of core-shell $Hf_xZr_{1-x}O_2$ nanoparticles

Theoretical predictions indicate that chemical stress can induce a ferroelectric o-III phase with related polar properties in spherical core-shell $HfO_2$ nanoparticles [14]. Similar effects may be expected in the oxygen-deficient core-shell $Hf_xZr_{1-x}O_2$ nanoparticles [27], where oxygen vacancies localized within a thin shell can induce strong chemical strains (elastic dipoles) [48, 49] inside the core of the particles.

To calculate the pyroelectric and electrocaloric properties of oxygen-deficient $Hf_xZr_{1-x}O_2$ nanoparticles, we consider densely packed spherical core-shell $Hf_{0.5}Zr_{0.5}O_2$ nanoparticles of the same radius $R$ and random orientation of crystallographic axes inside the particles. The cores are assumed to be defect-free, crystalline, and insulating (i.e., defect-free $Hf_{0.5}Zr_{0.5}O_2$). The core is surrounded by a thin shell, whose thickness $\Delta R$ is much smaller than the core radius $R_c$ (see **Fig. 1(a)**). The shell is assumed to be semiconducting due to the high concentration of oxygen vacancies (i.e., oxygen-deficient $Hf_{0.5}Zr_{0.5}O_2$ formed during annealing in a CO+$CO_2$ atmosphere), which can be readily ionized and therefore screen a polarization-related bound charge of the core and, consequently, reduce the associated depolarization field. Both neutral and ionized vacancies act as elastic dipoles, which induce strong chemical strains with a magnitude $w_s$. The chemical strain constrained by the elastic mismatch gives rise to elastic stress in the core [14]. The nanoparticles are assumed to be placed in a dielectric medium, whose effective dielectric permittivity $\varepsilon_{eff}$ depends on their concentration (see, e.g., Ref. [50] for details). Although free rotation of individual ferroelectric particles may occur, e.g., in liquid crystals, cohesive forces preclude such rotation in densely packed nanoparticles.

To describe the phase diagrams of the oxygen-deficient $Hf_xZr_{1-x}O_2$ nanoparticle, we use the LGD approach proposed in Ref. [14]. The LGD free energy functional of the $Hf_xZr_{1-x}O_2$ nanoparticle core, corresponding to the polar o-III phase, consists of the bulk energy density $f_{bulk}$, the electric energy $f_{el}$, and the surface energy $F_s$ [14]:

$$F_{o-III} = \int (f_{bulk} + f_{el}) dV + F_s, \quad f_{bulk} = f_{bq} + f_{tr} + f_{est} + f_{grad}. \tag{1}$$

The bulk energy density $f_{bulk}$ is the sum of the biquadratic energy $f_{bq}$; the trilinear coupling energy $f_{tr}$ of the polar, antipolar, and nonpolar order parameters; the elastic and striction energy contributions $f_{est}$, and the gradient energy of the order parameters $f_{grad}$. The electrostatic energy density $f_{el}$ has a conventional form, $f_{el} = -\vec{P}\vec{E}$, where $\vec{P}$ is an electric polarization and $\vec{E}$ is an electric field.

The energy density $f_{bq}$ is an expansion in even powers, and $f_{tr}$ is an expansion in odd powers of the dimensionless amplitudes $Q_{\Gamma 3}$, $Q_{Y2}$, and $Q_{Y4}$, corresponding to the polar, nonpolar, and antipolar phonon modes $\Gamma_{3-}$, $Y_{2+}$, and $Y_{4-}$, respectively (see Refs. [20, 21, 22] for details). The energies $f_{bq}$, $f_{tr}$, $f_{est}$, and $f_{grad}$ are listed in Ref. [14]. LGD parameters, determined from the DFT calculations [20] and piezoelectric response measurements [51], are listed in Ref. [16]. Assuming natural boundary conditions at the core-shell interface, namely $\frac{\partial Q_i}{\partial \vec{n}} = 0$, the surface energy $F_S$ does not contribute to the free energy in Eq. (1). Finally, the ferroelectric state of the core is stable if its free energy $F_{o-III}$ is smaller than the free energy of the m-phase, $F_m = \int f_m dV$, where $f_m$ is the energy density of the m-phase in bulk $Hf_xZr_{1-x}O_2$.

The form of the free energy functional is based on the work of Delodovici et al. [20], and Jung and Birol [21, 22], where the principal role of the trilinear coupling between the polar, antipolar, and nonpolar modes has been established. Following Lee et al. [52], the appearance of the polar o-III mode should be accompanied by the antipolar and nonpolar modes, because the trilinear coupling of the modes leads to the trigger-type ferroelectric transition to the o-III phase (see, e.g., Fig. 1 in Ref. [52]).

The total polarization $P_3$ directed along the "local" crystallographic (polar) axis "$z'$" of the individual nanoparticle (see **Fig. 1(a)**) is proportional to the amplitude $Q_{\Gamma 3}$ of the $\Gamma_{3-}$ mode, whereas the pyroelectric coefficient $\Pi$ and electrocaloric coefficient $\Sigma$ are proportional to its derivative [16, 20, 21]:

$$P_3 \approx P_0 Q_{\Gamma 3}, \tag{2a}$$

$$\Pi = -\left(\frac{\partial P_3}{\partial T}\right)_{E_3} \approx -P_0 \frac{\partial Q_{\Gamma 3}}{\partial T}, \tag{2b}$$

$$\Sigma = \left(\frac{\partial S}{\partial E_3}\right)_T = \left(\frac{\partial P_3}{\partial T}\right)_{E_3} = -\Pi, \tag{2c}$$

where the polarization amplitude $P_0$ is taken at zero Kelvin [20]. The value of $P_0$ is determined by the chemical strain produced by oxygen vacancies localized in the nanoparticle shell and can vary from 30 to 80 μC/cm$^2$ for high compressive strains [14]. $E_3$ is an external electric field. The equality $\left(\frac{\partial S}{\partial E_3}\right)_T = \left(\frac{\partial P_3}{\partial T}\right)_{E_3}$, Cursedwhich comes from Maxwell relations, provides the relation $\Sigma = -\Pi$. Here, $P_S$ denotes the spontaneous (zero-field) polarization, i.e., $P_S = P_3$ when $E_3 = 0$.

Note that the LGD parameters, determined in Ref. [16] from DFT calculations [20], depend on strain and corresponds to low temperatures, namely to zero Kelvin. The proposed temperature dependences of the spontaneous polarization, pyroelectric and electrocaloric coefficients are based on the experimental results presented in Refs. [31-33]. Considering the relaxor-type temperature decay of the remanent (spontaneous) polarization $P_S$ measured in $Hf_xZr_{1-x}O_2$ thin films (see, e.g., Fig. 5.2.2 in Ref. [33]), one can assume that the decay can be approximated by the exponential functions:

$$P_s \approx \begin{cases} P_0, & T < T_f, \\ P_0\left[1 - \exp\left(-\frac{T_a}{T-T_f}\right)\right], & T > T_f. \end{cases} \tag{3a}$$

$$\Pi = -\Sigma \approx \begin{cases} 0, & T < T_f, \\ \Pi_0 \exp\left(-\frac{T_a}{T-T_f}\right)\frac{T_a^2}{(T-T_f)^2}, & T > T_f. \end{cases} \tag{3b}$$

Here, $T_a$ is introduced as an activation energy parameter (in temperature units) associated with atomic thermal fluctuations that suppress and eventually destroy the long-range polar order. The so-called "freezing" temperature $T_f$ of electric dipolar order is strongly influenced by the concentration of oxygen vacancies (at fixed Zr content), which induce chemical strains that stabilize the FE o-III phase. The value of $T_f$ may also depend on the Zr content. The applicability of this phenomenological description is illustrated by the fit shown in Fig. S1. Note that the exponential function, namely $\exp\left(-\frac{T_a}{T-T_f}\right)$, resembles the Vogel-Fulcher law used to describe relaxor ferroelectrics [53] and dipolar glasses [54, 55]. The parameter $\Pi_0 = \frac{P_0}{T_a}$, introduced in Eq.(3b), is the magnitude of the pyroelectric/electrocaloric effect.

Using the same approximations and assumptions as those used in Eq. (3a), we assume that the polarization switching field $E_s$ is given by expression

$$E_s(T) \cong \begin{cases} E_0, & T < T_f, \\ E_0\left[1 - \exp\left(-\frac{T_a}{T-T_f}\right)\right]^3, & T > T_f. \end{cases} \tag{4}$$

The field decreases with increasing temperature above $T_f$ and vanishes at $T \gg T_f + T_a$ (see red curve in **Fig. 1(b)**). The field $E_s$ is analogous to the thermodynamic coercive field $E_c$ in a bulk material with a sharp ferroelectric-paraelectric phase transition. The field $E_0$ may be regarded as the thermodynamic coercive field at temperatures below $T_f$. Accordingly, $E_s$ represents the polarization switching field of a single-domain nanoparticle. The temperature dependence of $E_s$ can be explained using the LGD phenomenological approach. Since the coefficient $\alpha$, introduced in the Landau free energy density, $f_{bulk} = \frac{\alpha}{2}P_3^2 + \frac{\beta}{4}P_3^4 + ..$, is the only temperature-dependent coefficient in the LGD expansion, all temperature dependence of the coercive field $E_c \sim |\alpha|^{3/2}$ and spontaneous polarization $P_s \sim |\alpha|^{1/2}$ originates from $\alpha(T)$ in ferroelectrics undergoing a second order phase transition, as shown in the **Supplementary Material**. The assumed form of $\alpha(T)$ leads to the relations $E_s(T) \sim E_c(T)$ and $E_a(T) \sim \left(P_s(T)\right)^3$, which explain the temperature dependence of $E_s$ in Eq. (4).

Expressions (3)-(4) are valid for a single-domain $Hf_xZr_{1-x}O_2$ nanoparticle, whose spontaneous polarization is well-screened by ambient free charges accumulated at its surface, as well as for an ensemble of uniformly polarized single-domain nanoparticles, whose spontaneous polarizations are aligned due to strong pre-poling (e.g., during the preparation process). Corresponding temperature dependences of the spontaneous polarization $P_s$, pyroelectric coefficient $\Pi$, electrocaloric coefficient $\Sigma$, and polarization switching field $E_a$ are shown in **Fig. 1(b).** Hereinafter, we define the dimensionless reduced temperature by $\tau = \frac{T-T_f}{T_a}$, such that $\tau = 0$ corresponds to $T = T_f$ (see the dashed vertical line in **Fig. 1(b)**). The spontaneous polarization is constant ($P_0$) at $\tau < 0$ (corresponding to $T < T_f$); its gradual decrease is caused by the thermal activation of dipolar disorder, which begins at $\tau > 0$ (corresponding to $T > T_f$). The disorder dominates at $\tau \gg 1$ (which corresponds to $T \gg T_f + T_a$), where the spontaneous polarization becomes negligibly small. The temperature dependence of $P_s$ agrees qualitatively with typical experimental data for Si-doped hafnia thin films [33] (see **Fig. S2** in **Supplementary Materials**). The inset in **Fig. 1(b)** shows the polarized nanoparticles at $\tau < 0$ (corresponding to $T < T_f$).

The "zero-field" spatially averaged spontaneous polarization $\langle P_s \rangle$, pyroelectric coefficient $\langle \Pi \rangle$, and electrocaloric coefficient $\langle \Sigma \rangle$ are zero for the ensemble of $Hf_xZr_{1-x}O_2$ nanoparticles, because the spontaneous polarization vectors of individual particles are misaligned due to the random orientation of crystallographic axes. An ensemble of single-domain nanoparticles with zero average polarization is illustrated schematically in the upper-left inset of **Fig. 1(c)**. Electric fields significantly smaller than the coercive field cannot reorient the polarization inside the particles, because the considered densely packed particles cannot rotate due to cohesive forces.

For the considered case of a diffuse ferroelectric-paraelectric phase transition, described by Eqs. (3), the "field-induced" temperature dependences of the average values $\langle P_3 \rangle$, $\langle \Pi \rangle$, and $\langle \Sigma \rangle$ can be nonzero even at relatively small external electric fields $E_3 \ll E_0$ and at $T > T_f$, because a weak alignment of the particles' polarization (depicted by yellow arrows in the upper-right inset in **Fig. 1(c)**) occurs when $E_3 \geq E_s(T)$. Using the method of spatial averaging developed in Ref. [56], averaging over the random orientation of the crystallographic axes of the nanoparticles yields the following approximate expressions for $\langle P_3 \rangle$, $\langle \Pi \rangle$, and $\langle \Sigma \rangle$:

$$\langle P_3(\vec{E}) \rangle = \frac{1}{2}\int_0^{\pi} P_3(\vec{E})\sin\theta d\theta \approx \begin{cases} \frac{\eta}{4} P_0 \tanh\left(\frac{E_3}{E_0}\right), & T < T_f, \\ \frac{\eta}{4} P_0 \left[1 - \exp\left(-\frac{T_a}{T-T_f}\right)\right] \tanh\left(\frac{E_3}{E_s(T)}\right), & T > T_f, \end{cases} \tag{5a}$$

$$\langle \Pi \rangle = -\langle \Sigma \rangle \approx \begin{cases} 0, & T < T_f, \\ \frac{\eta}{4} P_0 \frac{\exp\left(-\frac{T_a}{T-T_f}\right)}{\left(T-T_f\right)^2} \left( \tanh\left(\frac{E_3}{E_s(T)}\right) - \frac{3 T_a E_3 \mathrm{sech}^2\left(\frac{E_3}{E_s(T)}\right)}{\left[1-\exp\left(-\frac{T_a}{T-T_f}\right)\right]^3} \right), & T > T_f. \end{cases} \tag{5b}$$

To derive Eqs. (5), we assumed that the polarization $P_3(\vec{E})$ satisfies the LGD-type equation inside individual nanoparticles, namely $\frac{\partial}{\partial P_3} f_{bulk}[P_3] = E_3 \cos\theta$, and performed the averaging over the polar angle $\theta$ in spherical coordinates (see **Supplement S1** in **Supplementary Materials** for mathematical details).

The positive constant $\eta$ is typically close to unity.

Corresponding temperature dependences of the field-induced average polarization $\langle P_3 \rangle$, pyroelectric and electrocaloric coefficients, $\langle \Pi \rangle$ and $\langle \Sigma \rangle$, calculated using Eqs. (5) for $E_3 = E_s$ at $\tau = 2$, are shown in **Fig. 1(c).** The average polarization $\langle P_3 \rangle$ is relatively small at $\tau < 0$ (because the applied field is much smaller than the switching field and so very few particles can reorient their polarization at $T < T_f$), reaches a diffuse maximum at $\tau = 2$ (since $E_3 = E_s$ at this temperature point $T = T_f + 2T_a$), and then decreases gradually and saturates at $\tau \gg 1$, because the spontaneous polarization of each particle decreases with increasing temperature and eventually dominates over the field-induced alignment at $T \gg T_f + T_a$ (see the green curve in **Fig. 1(c)**). The upper-left inset in **Fig. 1(c)** corresponds to the case $E_3 \ll E_s$, when the average spontaneous polarization is zero (see misaligned green arrows). Nanoparticles, whose polarization has switched (or become aligned) because $E_3 \gg E_s$, are depicted by yellow arrows in the upper-right inset in **Fig. 1(c)**. Since nanoparticles cannot rotate, their individual crystallographic axes are fixed in space. Thus, spontaneous polarization can be switched only between 2 directions, along or counter the polar axis. Of course, external field can induce some polarization directed along the field too, but the induced polarization has no bistable properties and is regarded small in comparison with the switchable spontaneous polarization. That is why yellow and green arrows in **Fig. 1(c)** still show the degree of misalignment in contrast to ideal alignment shown in **Fig. 1(a).**

The average electrocaloric coefficient $\langle \Sigma \rangle$ is zero at $\tau < 0$, reaches a maximum $\tau = 1$, decreases and equals to zero at $\tau = 2$ (since $\langle P_3 \rangle$ is maximal at this temperature point), then reaches a diffuse minimum at $\tau \approx 2.5$, and finally increases monotonically and tends to zero at $\tau \gg 1$ (see the red curve in **Fig. 1(c)**). Since $\langle \Pi \rangle = -\langle \Sigma \rangle$, the temperature behavior of the average pyroelectric coefficient $\langle \Pi \rangle$ is inverse to the behavior of $\langle \Sigma \rangle$ (see the blue curve in **Fig. 1(c)**).

Looking ahead, we note that the temperature dependences of $\langle P_3 \rangle$ and $\langle \Pi \rangle$ agree qualitatively with experimentally measured temperature dependences of electric charge $\Delta Q$, accumulated by pressed $Hf_xZr_{1-x}O_2$ nanoparticles, and its temperature derivative $\frac{\partial}{\partial T}\Delta Q$, respectively (experimental results will be presented in the next section).

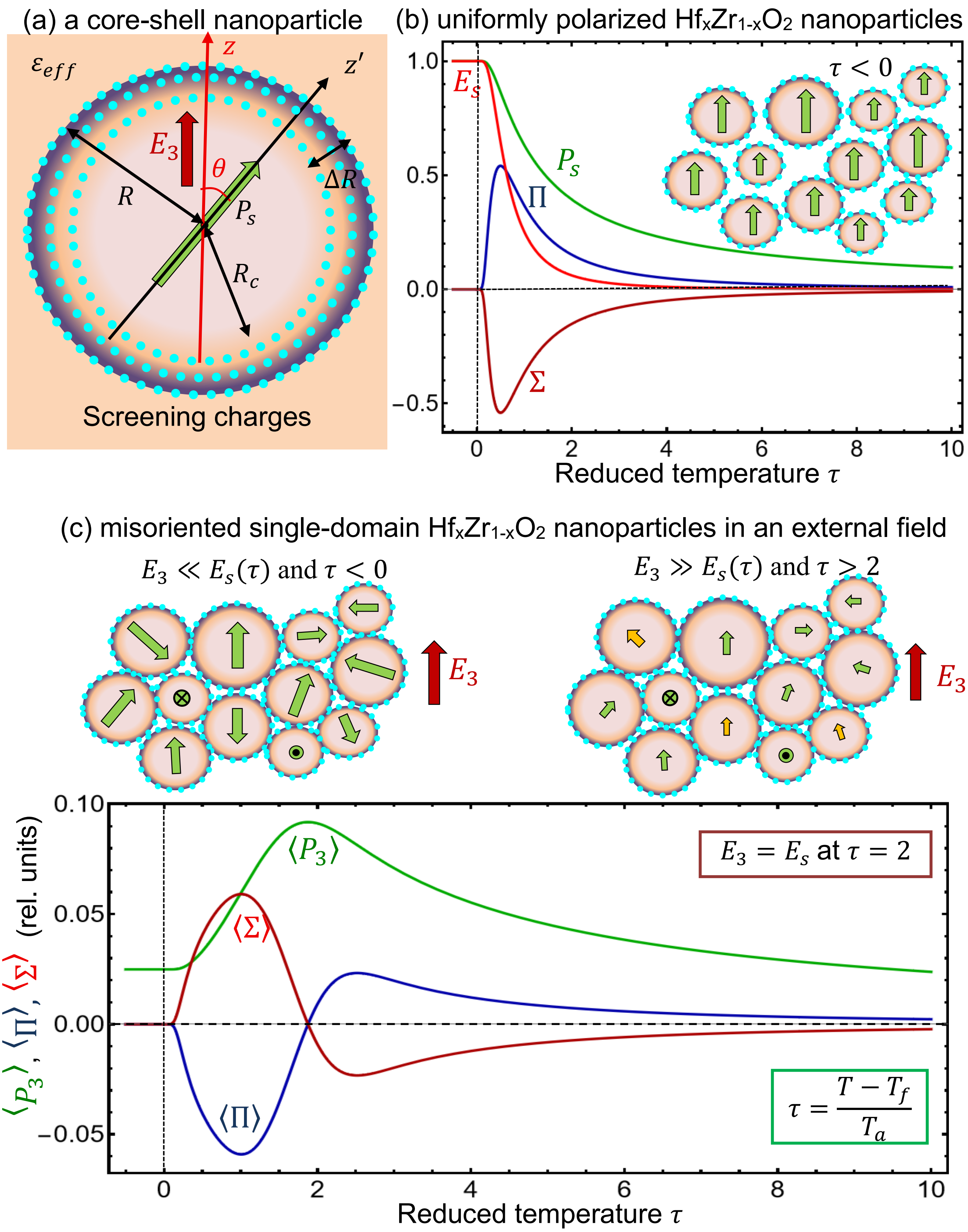

**Figure 1.** Comparison of two contrasting physical scenarios: a pre-poled (uniformly polarized) nanoparticle ensemble at zero external field and a misoriented nanoparticle ensemble subjected to an external electric field. **(a)** The cross-section of a spherical core-shell $Hf_xZr_{1-x}O_2$ nanoparticle: a defect-free core of radius $R_c$ is covered with a paraelectric shell of thickness $\Delta R$, which is filled with elastic defects (e.g., oxygen vacancies) and free screening charges (cyan circles). The green arrow shows the direction of spontaneous polarization $P_s$ in the core. The angle between the "global" direction "$z$" of the external field $E_3$ and the local polar axis "$z'$" is $\theta$. **(b)** Temperature dependences of the normalized spontaneous polarization $P_s/P_0$ (green curve), switching field $E_s/E_0$ (red curve), pyroelectric coefficient $\Pi/\Pi_0$ (blue curve), and electrocaloric coefficient $\Sigma/\Pi_0$ (dark-red curve). The dependences are calculated from Eq.(3) at zero external field ($E_3 = 0$). **(c)** Field-induced temperature dependences of the average spontaneous polarization $\langle P_3 \rangle / P_0$ (green curve), pyroelectric coefficient $\langle \Pi \rangle$ (blue curve), and electrocaloric coefficient $\langle \Sigma \rangle$ (red curve). The dependences are calculated from Eqs.(5) at $\eta = 1$ and external electric field $E_3 \approx E_s|_{\tau=2}$. The reduced temperature is $\tau = \frac{T-T_f}{T_a}$; the dashed vertical lines in parts (b) and (c) corresponds to $\tau = 0$. The inset in part (b) shows the polarized nanoparticles at $\tau < 0$. The upper-left inset in part (c) corresponds to the case $E_3 \ll E_s$, where the average spontaneous polarization is zero (see misaligned green arrows). The upper-right inset in part (c) corresponds to the case $E_3 \gg E_s$, where the spontaneous polarization of some nanoparticles has switched (see partially aligned yellow arrows).

## 3. Temperature-dependent charge accumulation in oxygen-deficient core-shell $Hf_{0.5}Zr_{0.5}O_2$ nanopowders

The studied oxygen-deficient $Hf_{0.5}Zr_{0.5}O_2$ nanoparticles were synthesized using solid-state organonitrate synthesis and annealed in a $CO$+$CO_2$ atmosphere to create oxygen vacancies (see preparation details in Refs. [36, 47]). Annealing in the $CO$+$CO_2$ atmosphere leads to the core-shell structure formation in the nanoparticles. The average size of the nanoparticles, determined from TEM and SEM studies, is 7 nm. The symmetry of the nanoparticles, determined from XRD analysis performed in Refs. [36, 47], consists of a mixture of predominantly orthorhombic phases (collectively accounting for more than 85 wt. %), including nonpolar o-I, antipolar o-II, and the desirable polar o-III phases, together with a small amount (less than 15 wt.%) of the monoclinic structure.

To study the dielectric response and accumulation of electric charge in the core-shell $Hf_{0.5}Zr_{0.5}O_2$ nanoparticles, the powder was cold-pressed at 5 MPa and room temperature into tablets and placed in a Teflon cylinder between two brass plungers, which provided uniaxial pressure and served as electrical contacts. The diameter of the samples was 4 mm, while their thickness and the distance between the contacts were (300 ± 20) µm. Measurements of the capacitance and resistance of the tablets were carried out using an RLC-meter (LCX200, Rohde & Schwarz) over the frequency range 4 Hz – 500 kHz (see details in Ref. [36]).

The porosity of the studied tablets was estimated as (20 – 30) % depending on the applied pressure. We consider the porosity in our theoretical calculations by using the EMA (see details in **Supplement S2** of the **Supplementary Materials**). Within the EMA, the pressed nanopowder is considered as a binary mixture of quasi-spherical, densely packed oxygen-deficient core-shell $Hf_xZr_{1-x}O_2$ nanoparticles (with the volume fraction μ ≈ 0.75) and air (with the volume fraction 1−μ ≈ 0.25). This allows us to separate the response of the core-shell nanoparticles, as shown in **Fig. S2** of the **Supplementary Materials.**

To characterize the dynamic electrical response, we performed real time measurements of charge-voltage $Q(V)$ and current-voltage $I(V)$ dependences in the alternating current (AC) mode at room temperature. Charge-voltage and current-voltage loops for the $Hf_{0.5}Zr_{0.5}O_2$ nanopowder sample obtained under bipolar AC triangle driving voltage, as well as current-voltage curves obtained under a unipolar saw-tooth driving voltage, are presented in **Fig. 2(a)** and **2(b)**, respectively. The bipolar $Q(V)$ loop has a shape resembling that of a relaxor ferroelectric [53-55], with a weak but noticeable tendency toward saturation as the voltage amplitude increases. The shape of the bipolar $I(V)$ loop is defined by the contributions of an effective capacitance $C$ and resistance $R$ connected in parallel. The effective capacitance $C \approx 5$ pF and resistance $R \approx 0.5$ GΩ were estimated by fitting the measured $Q(V)$ and $I(V)$ loops to an equivalent parallel RC circuit. The fitting produces parallelogram-like $Q(V)$ and $I(V)$ loops whose shape and sizes resemble those of the measured $Q(V)$ and $I(V)$ loops, except for the smooth angles (compare dotted parallelogram with the measured loop in **Fig. 2(b)**). The smoothed angles of the bipolar parallelogram-like $Q(V)$ and $I(V)$ loops, as well as the saturated behavior of the unipolar $I(V)$ curve, can be attributed to the MWS-type polarization emerging due to the accumulation of space charge within the shells and interparticle spaces.

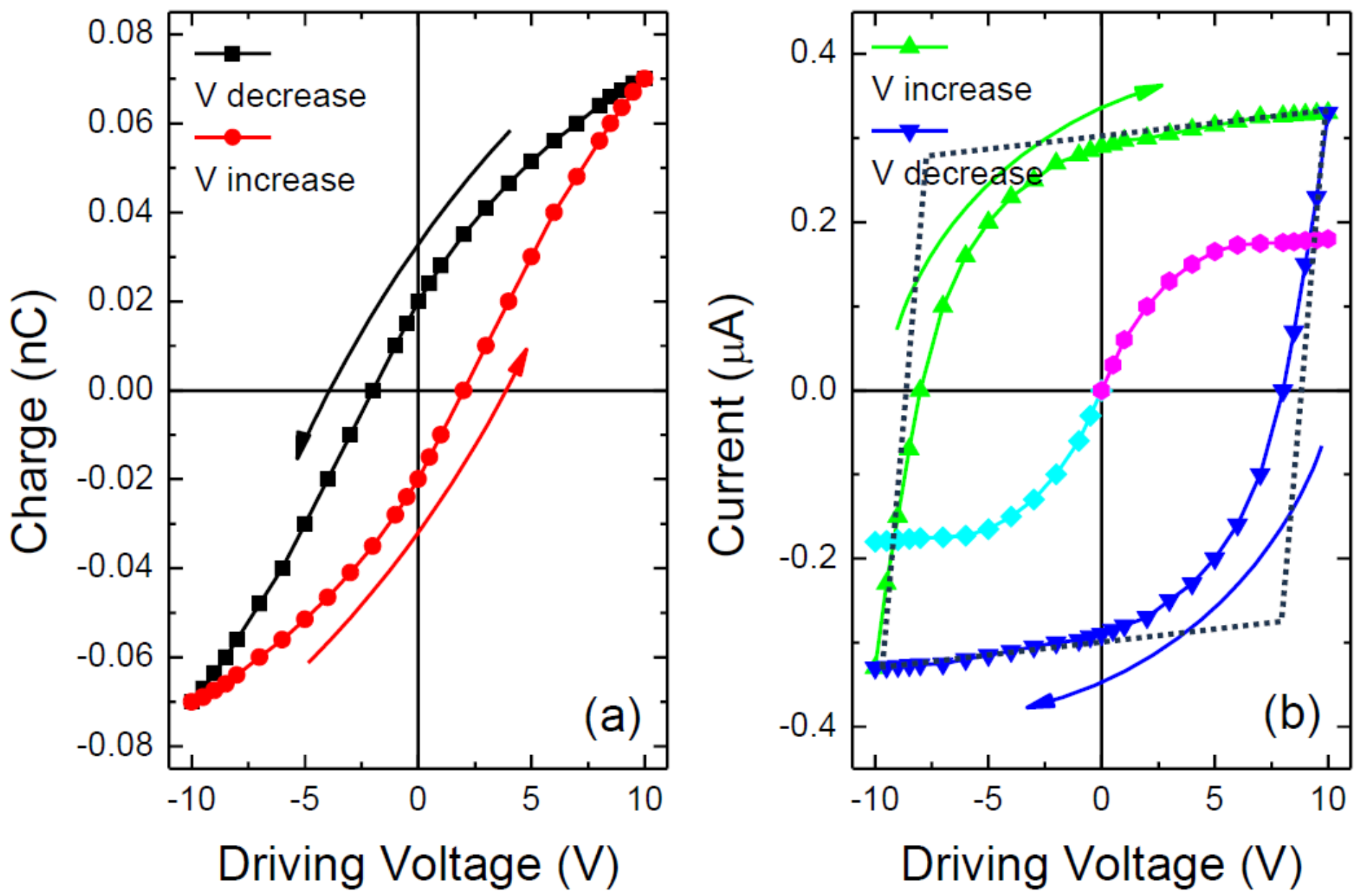

**Figure 2.** The charge-voltage loop $Q(V)$ (red and black rectangles) **(a)** and current-voltage loop $I(V)$ (green and blue triangles) **(b)** measured under a bipolar AC triangular voltage. The current-voltage curve $I(V)$ (magenta and cyan circles) **(b)** measured under a unipolar saw-tooth driving voltage. Different colors of symbols correspond to different directions (increase or decrease) of the driving voltage amplitude, also indicated by arrows. All measurements were performed on the $Hf_{0.5}Zr_{0.5}O_2$ tablet at 1 kHz. Dotted parallelogram in the part (b) corresponds to the RC-circuit response.

Next, we performed measurements of the temperature dependence of the electric charge accumulated in the pressed $Hf_{0.5}Zr_{0.5}O_2$ tablets during direct current (DC) flow. For this purpose, we measured the time dependence of the charging current at a constant applied voltage of 2V across the sample at different temperatures $T$ (see details in Ref. [36]). To calculate the accumulated charge, we used exponential functions to fit the measured current:

$$J(t) = J_0 + \sum_{i=1}^{N} A_i \exp\left(-\frac{t}{\tau_i}\right). \quad (6a)$$

The corresponding accumulated charge is given by:

$$\Delta Q = \int_0^\infty dt\,(J(t) - J_0) = \sum_{i=1}^{N} A_i \int_0^\infty dt\,\exp\left(-\frac{t}{\tau_i}\right) = \sum_{i=1}^{N} A_i \tau_i. \quad (6b)$$

The details of the determination of $\Delta Q$ are given in Ref. [36]. The total charge $\Delta Q$ accumulated in the pressed core-shell $Hf_{0.5}Zr_{0.5}O_2$ nanopowder as a function of temperature is shown in **Fig. 3(a)**. It is seen from **Fig. 3(a)** that the accumulated electric charge is maximal in the temperature range 60 – 70°C; it decreases rapidly above 70°C and becomes negligible above 105°C. The temperature dependence of $\Delta Q$ resembles qualitatively the temperature dependence of the average polarization $\langle P_3 \rangle$ at $T > T_f$, shown in **Fig. 1(c)**.

Notably, the total discharge time after the voltage removal is more than 20 minutes, which may be indicative of pronounced electret-like properties of the samples [36]. The effective charge density, $\Delta Q/S$, can be estimated using the total contact area, $S = 12.56$ mm$^2$, yielding a maximal value of 254 μC/cm$^2$. It is unlikely that such a large charge density originates solely from ferroelectric polarization (i.e., bound charge). Instead, it likely results from the cumulative contributions of the strain-induced ferroelectricity in the cores of $Hf_{0.5}Zr_{0.5}O_2$ nanoparticles and the IBLC and SBLC effects in their oxygen-deficient shells. To disentangle these contributions, permittivity measurements at a frequency higher than 1 MHz may be in order. At the same time a large permittivity at low frequencies, that disappear well below 1 MHz, is characteristic to the MWS-type polarization originating due to the IBLC and SBLC effects [36]. A detailed separation of these contributions, however, is beyond the scope of the present study.

The temperature derivative of the accumulated charge, $\frac{\partial}{\partial T}\Delta Q$ (an analog of the pyroelectric coefficient), is shown in **Fig. 3(b)**. The derivative is positive and decreases monotonically over the temperature range 20 – 62°C becoming zero near 62°C. It then becomes negative, reaches a pronounced minimum at 89°C, and subsequently increases, approaching zero above 105°C. The temperature dependence of $\frac{\partial}{\partial T}\Delta Q$ qualitatively resembles that of the average pyroelectric coefficient $\langle \Pi \rangle$ at $T > 2T_f$, shown in **Fig. 1(c)**. This resemblance means that the contribution of ferroelectric dipoles to the measured $\frac{\partial}{\partial T}\Delta Q$ can be significant. Based on the similarity in the shapes of the curves in **Fig. 3(b)** and **Fig. 1(c)**, but recognizing that the magnitudes of $\Delta Q/S$ and $P_s$ cannot be directly compared, we may assume that the ferroelectric dipoles create potential barriers that facilitate the accumulation of space charge through screening mechanisms. Under this assumption, the temperature dependence of $\Delta Q(T)$ is determined by both the spontaneous polarization $P_s(T)$ and the MWS-type polarization. The barrier height can be proportional to $P_s$, with the proportionality coefficient determined by the degree of screening. However, the relative contribution of $P_s$ to $\Delta Q(T)$ cannot be estimated from the performed experiments.

The qualitative agreement between the experimentally measured $\Delta Q$ and the calculated average polarization $\langle P_3 \rangle$, as well as between the experimentally determined $\frac{\partial}{\partial T}\Delta Q$ and the calculated average pyroelectric and electrocaloric coefficients, $\langle \Pi \rangle$ and $\langle \Sigma \rangle$, probably indicates the existence of ferroelectric properties and, consequently, pyroelectric and electrocaloric properties in the oxygen-deficient core-shell $Hf_{0.5}Zr_{0.5}O_2$ nanoparticles.

Factors proportional to the temperature derivative of the accumulated charge, whose dependence on $\frac{\partial}{\partial T}\Delta Q$ and other parameters is the same as for the pyroelectric figures of merit [57], can be introduced as

$$\Phi_I = \frac{1}{C_p S}\left|\frac{\partial}{\partial T}\Delta Q\right|, \qquad \Phi_V = \frac{1}{C_p \varepsilon_0 \varepsilon_{NP} S}\left|\frac{\partial}{\partial T}\Delta Q\right|, \tag{7}$$

where $C_p = 2.72 \cdot 10^6$ J/(m$^3$K) is the specific heat per unit volume of the $Hf_{0.5}Zr_{0.5}O_2$ [58], $S = 12.56$ mm$^2$ is the area of each electrode, $\varepsilon_0 = 8.85 \cdot 10^{-12}$ F/m is the permittivity of free space, and $\varepsilon_{NP}$ is the quasi-static dielectric permittivity of the $Hf_{0.5}Zr_{0.5}O_2$ nanoparticles (shown by filled symbols in **Fig. S2**).

Factors $\Phi_I$ and $\Phi_V$ (analogs of the pyroelectric figures of merit) are shown in **Fig. 3(c)** and **3(d)**, respectively. Both factors are zero near 62 – 64°C, have a pronounced maximum near 90°C, decrease sharply at higher temperatures, and become zero above 105°C. The factors $\Phi_I$ and $\Phi_V$, which are positive and differ by the temperature-dependent factor $\frac{1}{\varepsilon_0 \varepsilon_{NP}}$, characterize electric charge accumulation in two different regimes: the short-circuited regime (subscript "$I$") and the open-circuited regime (subscript "$V$"). The factor $\frac{1}{\varepsilon_0 \varepsilon_{NP}}$ reflects the effective capacitance of the material in the open-circuited regime. These two regimes are the primary conditions for characterization of pyroelectric materials and operation of pyroelectric elements, which generate electric charge in response to temperature changes [57].

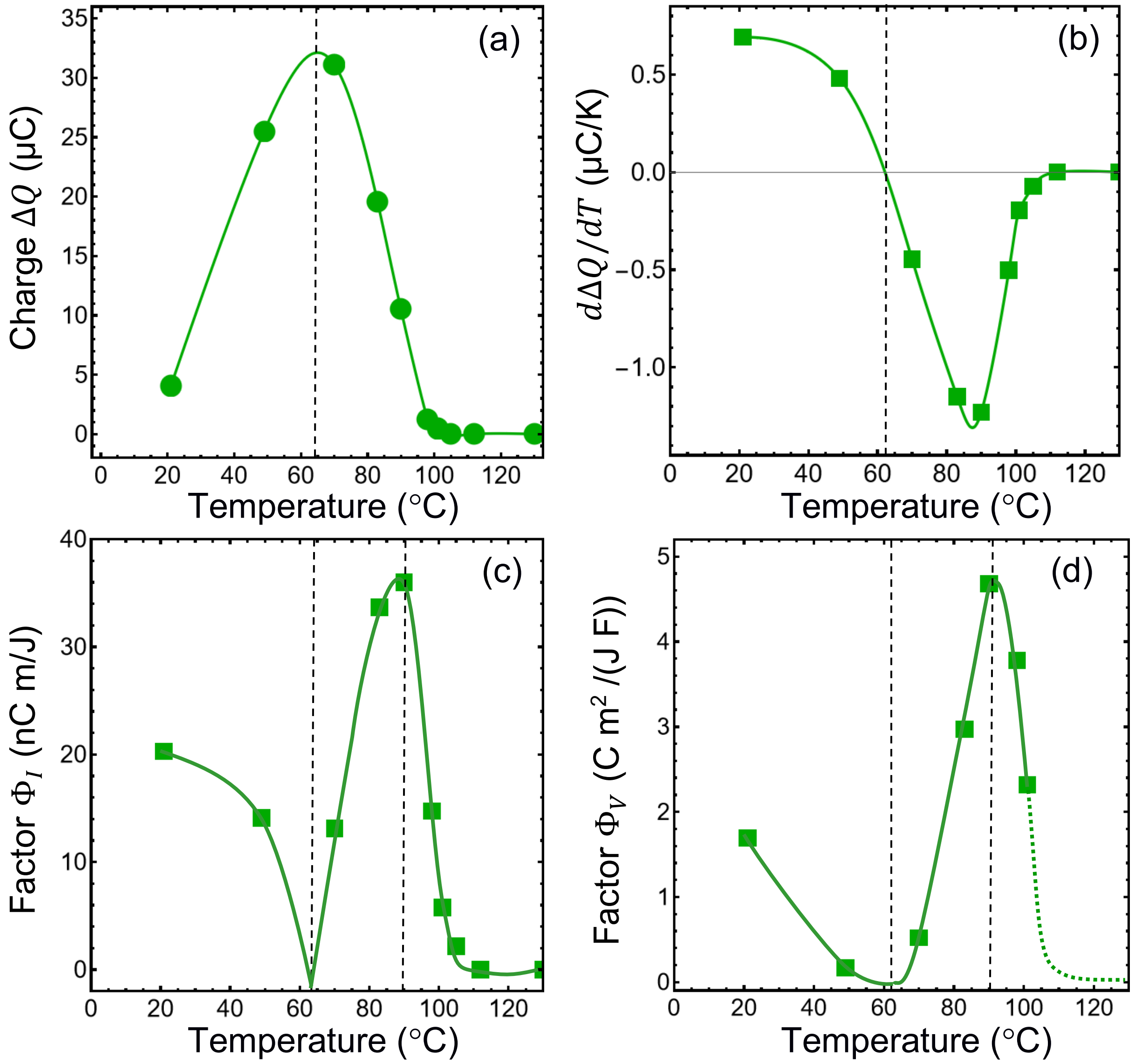

**Figure 3**. **(a)** Temperature dependence of the electric charge accumulated in pressed $Hf_{0.5}Zr_{0.5}O_2$ nanopowders. **(b)** Temperature derivative of accumulated charge, $\frac{\partial Q}{\partial T}$ (an analog of pyroelectric coefficient, $\langle \Pi \rangle$). Analogs of the pyroelectric figures of merit: $\Phi_I$ **(c)** and $\Phi_V$ **(d)**. The dotted green curve in **(d)** corresponds to an interpolation over the temperature range in which the dielectric permittivity was not measured (see **Fig. S2** for detail).

## 4. Conclusions

Using the Landau-Ginzburg-Devonshire free energy functional with trilinear and biquadratic couplings of polar, antipolar, and nonpolar order parameters, we analyzed the pyroelectric and electrocaloric properties of densely packed ensemble of spherical core-shell $Hf_xZr_{1-x}O_2$ nanoparticles with different orientations of the crystallographic axes. The "zero-field" spatially averaged polarization $\langle P_3 \rangle$, pyroelectric coefficient $\langle \Pi \rangle$, and electrocaloric coefficient $\langle \Sigma \rangle$ are zero because of the random orientation of crystallographic axes inside the particles. For the considered case of a diffuse ferroelectric-paraelectric phase transition, the field-induced temperature dependences of the average values $\langle P_3 \rangle$, $\langle \Pi \rangle$, and $\langle \Sigma \rangle$ can be nonzero at temperatures $T \gg T_f + T_a$ ($T_f$ is the freezing temperature and $T_a$ is the activation temperature) even in relatively small external electric fields. For small electric fields, the average polarization $\langle P_3 \rangle$ is negligibly small at temperatures $T < T_f$, reaches a broad maximum at $T = T_m$, where $T_m$ is determined by the field magnitude, and then decreases slowly, approaching a small value for $T \gg T_f + T_a$. The average pyroelectric coefficient $\langle \Pi \rangle$ is zero at $T < T_f$, reaches a minimum, increases to zero at $T = T_m$, reaches a broad maximum before gradually vanishing for $T \gg T_f + T_a$. The temperature dependence of the electrocaloric coefficient $\langle \Sigma \rangle$ mirrors that of the pyroelectric coefficient $\langle \Pi \rangle$, with an opposite sign, since $\langle \Sigma \rangle = -\langle \Pi \rangle$ according to the thermodynamic Maxwell relations.

Complementary to theoretical calculations, we performed temperature dependent measurements of the electric charge, $\Delta Q$, accumulated in pressed powders consisting of oxygen-deficient core-shell $Hf_{0.5}Zr_{0.5}O_2$ nanoparticles with an average size of 7 nm. Comparison with theoretical results allowed us to explain the experimentally observed temperature behavior of $\Delta Q$ and its temperature derivative $\frac{\partial}{\partial T}\Delta Q$, in the pressed tablets of $Hf_{0.5}Zr_{0.5}O_2$. Specifically, the calculated temperature dependences of $\langle P_3 \rangle$, $\langle \Sigma \rangle$, and $\langle \Pi \rangle$ agree qualitatively with experimentally measured temperature dependences of electric charge $\Delta Q$ and its temperature derivative $\frac{\partial}{\partial T}\Delta Q$, respectively. The qualitative agreement between experimentally measured $\Delta Q$ and calculated average polarization $\langle P_3 \rangle$, as well as between experimentally measured $\frac{\partial}{\partial T}\Delta Q$ and calculated average pyroelectric and electrocaloric coefficients $\langle \Pi \rangle$ and $\langle \Sigma \rangle$, suggests that ferroelectric dipoles contribute significantly to the measured charge accumulation and its temperature dependence. This interpretation is consistent with the existence of ferroelectric, pyroelectric, and electrocaloric properties in the oxygen-deficient core-shell $Hf_{0.5}Zr_{0.5}O_2$ nanoparticles. Thus, we believe these results can pave the way for the development of CMOS-compatible oxygen-deficient core-shell $Hf_xZr_{1-x}O_2$ nanoparticles for pyroelectric and electrocaloric applications.

**Authors' contribution.** A.N.M. performed analytical calculations, prepared corresponding figures, and wrote the manuscript draft. O.S.P. and N.V.M. performed the measurements and analyzed the results. E.A.E. performed numerical calculations, processed the experimental results, and prepared corresponding figures. D.R.E. analyzed the results and made all improvements in the manuscript.

**Data availability.** The data are available on reasonable request from the authors.

**Acknowledgments.** A.N.M. primary acknowledges the EOARD project 9IOE063 (related STCU partner project is P751c), and also acknowledges (analysis part) the National Research Foundation of Ukraine (grant

N 2023.03/0132 “Manyfold-degenerated metastable states of spontaneous polarization in nanoferroics: theory, experiment and perspectives for digital nanoelectronics”, and (discussion part) the Horizon Europe Framework Programme (HORIZON-TMA-MSCA-SE), project № 101131229, Piezoelectricity in 2D-materials: materials, modeling, and applications (PIEZO-2D). O.S.P. work is funded the National Research Foundation of Ukraine (grant N 2023.03/0132 “Manyfold-degenerated metastable states of spontaneous polarization in nanoferroics: theory, experiment and perspectives for digital nanoelectronics”. N.V.M. acknowledges the Target Program of the National Academy of Sciences of Ukraine, Project No. 5.8/26-П “Energy-saving and environmentally friendly nanoscale ferroics for the development of sensorics, nanoelectronics and spintronics”. The work of E.A.E. is funded by the National Research Foundation of Ukraine (grant N 2023.03/0127 “Silicon-compatible ferroelectric nanocomposites for electronics and sensors”). Results of theoretical modelling were visualized in Mathematica 14.0 [59].

# SUPPLEMENTARY MATERIAL TO

### Appendix S1. Mathematical details of polarization averaging

Polarization $P_3(\vec{E})$ satisfies the LGD-type equation inside individual nanoparticles, namely

$$\frac{\partial f_{bulk}}{\partial P_3} = E_3\cos\theta. \qquad \text{(S.1a)}$$

For a simplified model system that includes both order-disorder and displacive contributions to the polarization [60], the derivative $\frac{\partial f_{bulk}}{\partial P_3}$ can be approximated as:

$$\frac{\partial}{\partial P_3} f_{bulk} \approx \alpha(T)P_3 + \beta P_3^3 + \gamma P_3^5 + \delta P_3^7 + \cdots + E_s \text{arctanh}\left(\frac{P_3}{P_s}\right), \qquad \text{(S.1b)}$$

where $\alpha(T)$, $\beta$, $\gamma$, and $\delta$ are the LGD expansion coefficients; $E_s \sim \frac{k_B T}{P_s V}$ is the switching field determined by thermal activation mechanism, and $P_s$ is the spontaneous polarization.

Equation (S.1a) has a formal solution,

$$P_3 = P_s \tanh\left[\frac{E_3\cos\theta - \alpha(T)P_3 - \beta P_3^3 - \gamma P_3^5 - \delta P_3^7}{E_s}\right]. \qquad \text{(S.2)}$$

Under the condition $\beta > 0$ assuming negligibly contributions from the fifth and higher powers of the polarization, we calculate the average value:

$$\langle P_3(\vec{E})\rangle = \frac{P_s}{2}\int_0^{\pi} \tanh\left[\frac{E_3\cos\theta - \alpha(T)P_3 - \beta P_3^3}{E_s}\right] \sin\theta d\theta = \frac{P_s}{2E_3}\int_{E_3}^{-E_3} \tanh\left[\frac{E - \alpha(T)P_E - \beta P_E^3}{E_s}\right] dE, \qquad \text{(S.3)}$$

where $P_E$ satisfies the equation $\alpha(T)P_E + \beta P_E^3 + E_a \text{arctanh}\left(\frac{P_E}{P_s}\right) = E$. At very large external fields, $E_3 \gg E_s$, the polarization $\vec{P}$ of those nanoparticles whose polar axis has a nonzero projection onto the external field direction (z-axis) and satisfies $E_3\cos\theta > E_s$, aligns to minimize the dot product $\vec{P}\vec{E}$ and approaches a maximal value $P_m$ that cannot overcome $P_s$ in accordance with Eq. (S.2). The positive sign of the dot product $\langle\vec{P}\rangle\vec{E}$ corresponds to the minimum electrostatic energy of the system (see the upper-right inset in **Fig. 1(c)**, main text). For the case of large fields, averaging over the polarization directions leads to the expression

$$\langle P_3(\vec{E})\rangle \approx \frac{E_a}{8E_3}P_s \ln\left[\cosh\left(\frac{E-\alpha(T)P_m-\beta P_m^3}{E_s}\right)\right]\Big|_{E_3}^{-E_3} \approx \frac{\alpha(T)P_m+\beta P_m^3}{4E_3}P_s\tanh\left[\frac{E_3}{E_s}\right] \approx \frac{\eta}{4}P_s\tanh\left[\frac{E_3}{E_s}\right], \quad \text{(S.4)}$$

The factor $\frac{1}{4}$ originates from averaging over all polar-axis directions in spherical coordinates. We further assume that $\alpha P_m + \beta P_m^3 \cong \eta E_3$ due to the alignment of the elementary dipoles, where the factor $\eta$ is close to unity.

Under the condition $\beta > 0$, negligibly small contribution from the fifth and higher powers of the polarization, and weak thermal disorder, the expressions for the spontaneous polarization $P_s$ and coercive field $E_c$ are well-known. Specifically, $P_s(T) \cong \sqrt{-\frac{\alpha(T)}{\beta}}$ and $E_c(T) \cong \frac{2}{3\sqrt{3}}\alpha(T)\sqrt{-\frac{\alpha(T)}{\beta}}$. Hereinafter we use the approximation $\alpha(T) = \alpha_0$ at $T < T_f$, and $\alpha(T) = \alpha_0\left[1 - C_f\exp\left(-\frac{T_a}{T-T_f}\right)\right]^2$ for $T > T_f$. Using the approximation one can derive the dependences:

$$P_s \approx \begin{cases} P_0, & T < T_f, \\ P_0\left[1 - C_f\exp\left(-\frac{T_a}{T-T_f}\right)\right], & T > T_f. \end{cases} \quad \text{(S.5)}$$

Here, $T_a$ is the activation energy of atomic thermal fluctuations (in temperature units), which suppress and eventually destroy the long-range polar order, and $T_f$ is the "freezing" temperature of electric dipolar order. In general, the dimensionless constant $C_f$ is temperature-independent and equal to unity (see e.g., parameters in **Fig. S1**). However, in some cases, the remanent polarization increases slightly with increasing temperature, and the rate of increase depends on the amplitude and frequency of the applied electric field (see, e.g., Fig. 1(b) in Ref. [61]). This behavior can be described by Eq. (S.5) if the previously constant parameter $C_f$ is replaced by the temperature-dependent function $C_f(T)$, where $C_f(T)$ is small and negative at temperatures $T$ slightly higher than $T_f$. For $T \gg T_f$, $C_f(T)$ becomes positive and tends to unity. Although Eq. (S.5) can be extended in this way to describe the behavior reported in Refs. [34, 62, 63], the remainder of the present work considers the original temperature-independent case ($C_f = 1$).

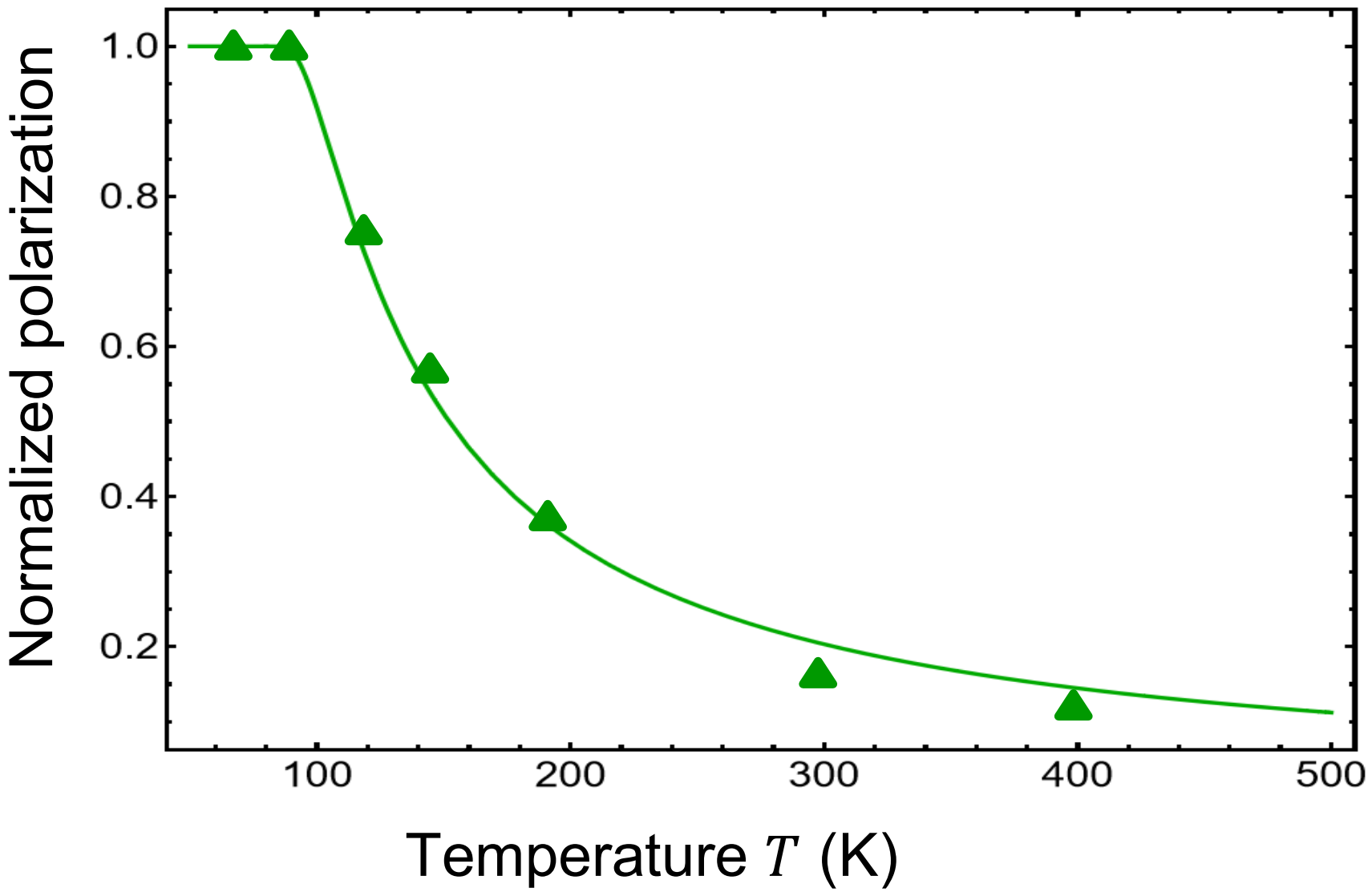

**FIGURE S1.** Typical temperature dependence of the remanent polarization in Si-doped $HfO_2$ thin films. Experimental results (shown by triangles) for $HfO_2$ thin films doped with 5.6 mol. % Si are taken from Ref. [62]. The solid curve shows the fit of expression (S.5) to the experimental data using $C_f = 1$, $T_a = 50$ K, and $T_f = 80$ K.

## Appendix S2. Mathematical details of effective dielectric permittivity calculations

The colossal effective dielectric permittivity $\varepsilon^*_{eff}$ was observed experimentally [64] for pressed nanopowders (see **Fig. S2**, open symbols). Such behavior of $\varepsilon^*_{eff}$ may be attributed to the interplay of dipolar polarization and IBLC and SBLC effects [36]. Within the EMA, this interplay is described by an algebraic equation for determining the complex dielectric permittivity $\varepsilon^*_{eff}$ of the effective medium [36, 65]:

$$(1-\mu)\frac{\varepsilon^*_{eff}-\varepsilon^*_a}{(1-\eta_{NP})\varepsilon^*_{eff}+\eta_{NP}\,\varepsilon^*_a}+\mu\frac{\varepsilon^*_{eff}-\varepsilon^*_{NP}}{(1-\eta_{NP})\varepsilon^*_{eff}+\eta_{NP}\,\varepsilon^*_{NP}}=0. \qquad \text{(S.6)}$$

Here, $\varepsilon^*_{NP}$ is the complex relative permittivity of the polarized nanoparticles "*NP*", which are assumed to be monodisperse and uniformly polarized. The real part of $\varepsilon^*_{NP}$ corresponds to the dielectric (polarization) response, and the imaginary part of $\varepsilon^*_{NP}$ accounts for dielectric losses due to finite ionic conductivity. These losses give rise to the MWS-type effects. $\varepsilon^*_a$ is the complex relative permittivity of the surrounding media "$a$", where $\mathrm{Re}[\varepsilon^*_a] \gg \mathrm{Im}[\varepsilon^*_a]$. The values of $\mu$ and $1-\mu$ are the relative volume fractions of the components "*NP*" and "$a$", respectively. The maximal value $\mu_{max}$ is limited by the dense packing of the particles. In particular, $\mu_{max} \approx 0.75$ for densely packed nanospheres of a uniform radius. The function $n_{NP}$ is the depolarization field factor of the polarized nanoparticles, which is determined by their shape and polarization orientation; $0 \le n_{NP} \le 1$. Analytical expressions for $n_{NP}$ exist for uniformly polarized nanoparticles of ellipsoidal shape, including nanowires, nanospheres, and nanodisks [66]. Screening charges present in the nanoparticle shell weaken the depolarization field inside the particle, and the corresponding decrease of the "effective" depolarization factor $\eta_{eff}$ can be roughly estimated as $\frac{\eta_{NP}}{1+(R_c/\lambda_{eff})}$. For instance, $\eta_{eff} < 0.1\eta_{NP}$ for $\lambda_{eff} < 1$ nm and $R_c$ >10 nm.

Assuming that $\eta_{NP} \to 0$, which corresponds to well-screened ($\lambda_{eff} \leq 0.1$ nm) nanoparticles and/or to the case where they form percolation clusters between the electrodes, we obtain from Eq. (S.5) that $\varepsilon^*_{eff} = \varepsilon^*_a(1-\mu) + \mu\varepsilon^*_{NP}$. Assuming that $\eta_{NP} = 1/3$, which corresponds to spatially isolated, weakly screened ($\lambda_{eff} \gg 1$ nm) spherical nanoparticles, the expression for $\varepsilon^*_{NP}$ has a relatively simple form:

$$\varepsilon^*_{NP} = \varepsilon^*_{eff}\frac{\mu\left[(1-\eta_{NP})\varepsilon^*_{eff}+\eta_{NP}\,\varepsilon^*_a\right]+(1-\eta_{NP})(1-\mu)\left(\varepsilon^*_{eff}-\varepsilon^*_a\right)}{\mu\left[(1-\eta_{NP})\varepsilon^*_{eff}+\eta_{NP}\,\varepsilon^*_a\right]-\eta_{NP}\,(1-\mu)\left(\varepsilon^*_{eff}-\varepsilon^*_a\right)} = \begin{cases} \frac{\varepsilon^*_{eff}}{\mu} - \varepsilon^*_a\frac{1-\mu}{\mu}, & \eta_{NP} \to 0, \\ \varepsilon^*_{eff}\frac{\mu\left[2\varepsilon^*_{eff}+\varepsilon^*_a\right]+2(1-\mu)\left(\varepsilon^*_{eff}-\varepsilon^*_a\right)}{\mu\left[2\varepsilon^*_{eff}+\varepsilon^*_a\right]-(1-\mu)\left(\varepsilon^*_{eff}-\varepsilon^*_a\right)}, & \eta_{NP} = \frac{1}{3}. \end{cases} \tag{S.7}$$

Experimentally measured [36] temperature dependences of the effective dielectric permittivity of pressed core-shell $Hf_{0.5}Zr_{0.5}O_2$ nanopowders (open symbols) and the dielectric permittivity of the $Hf_{0.5}Zr_{0.5}O_2$ nanoparticles deconvoluted using Eq. (S.6) at $\eta_{NP} = 1/3$ and $\mu = 0.75$ (filled symbols) are shown in **Fig. S2**. Circles and triangles show, respectively, the real and imaginary parts of the dielectric permittivity at test-signal frequencies of 4 Hz (**Fig. S2(a)**) and 500 kHz (**Fig. S2(b)**).

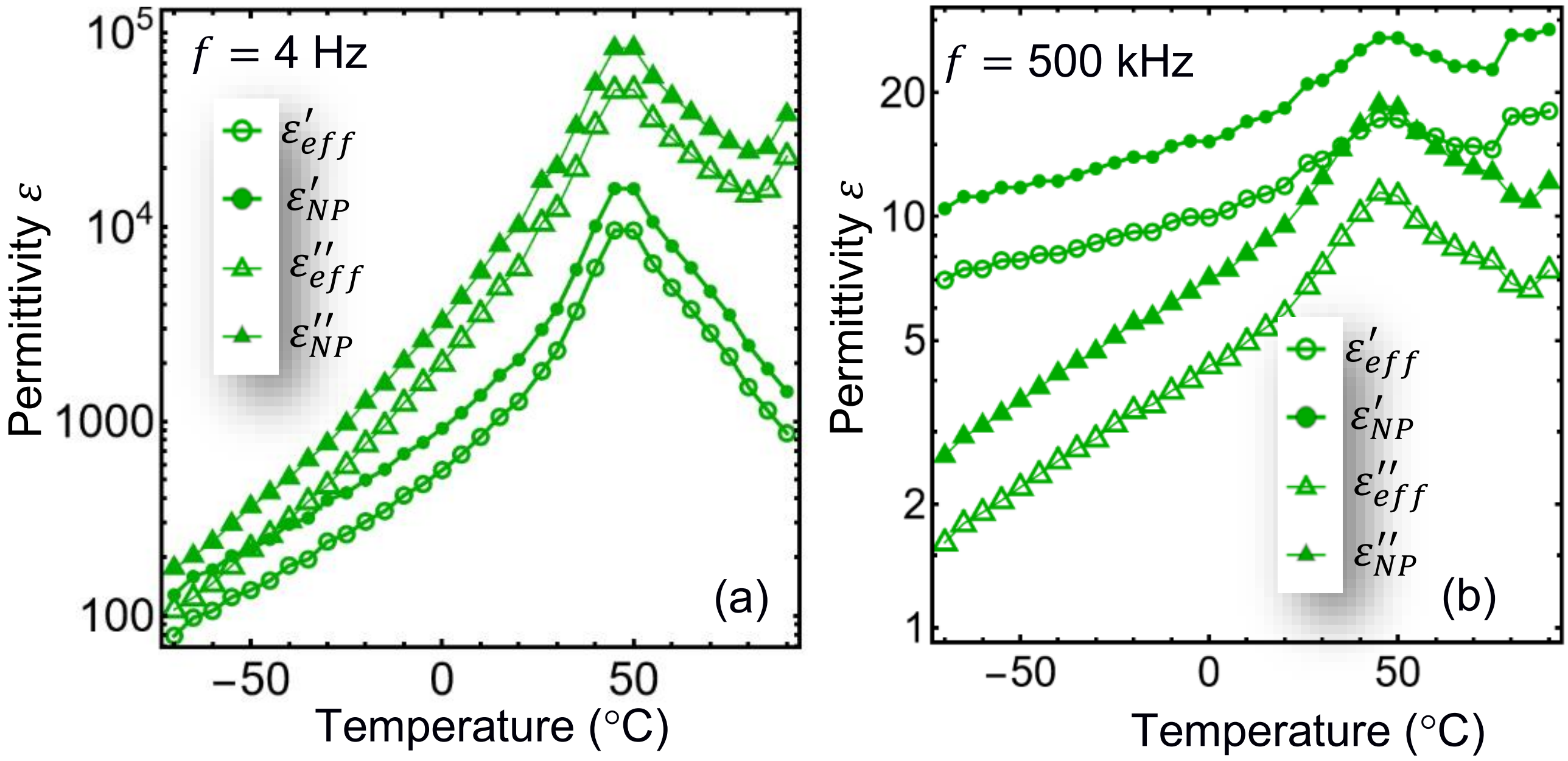

**FIGURE S2**. Temperature dependences of the effective dielectric permittivity $\varepsilon^*_{eff}$ of pressed core-shell $Hf_{0.5}Zr_{0.5}O_2$ nanopowders (open symbols) and of the dielectric permittivity of $Hf_{0.5}Zr_{0.5}O_2$ nanoparticles $\varepsilon^*_{NP}$ (filled symbols) "deconvoluted" using Eq. (S.7) at $\eta_{NP} = 1/3$ and $\mu = 0.75$. Circles and triangles show, respectively, the real and imaginary parts of the dielectric permittivity at test-signal frequencies 4 Hz **(a)** and 500 kHz (**b**). Experimental data (open symbols) were measured in Ref. [36].